\documentclass[sigconf]{acmart}
\usepackage{setspace}
\usepackage{censor}
\usepackage{balance}

\AtBeginDocument{%
  }

\setcopyright{none}
\acmYear{2018}
\acmDOI{XXXXXXX.XXXXXXX}
\acmConference[LIMITS '26]{12th Workshop on Computing within Limits}{June 23--25,
  2026}{Online}
\acmISBN{}

\begin{document}

\title{Narratives That Limit the Possible: Interrupting Narrative Closure in Computing Practice}


\author{Samuel Mann}
\affiliation{%
  \institution{Otago Polytechnic}
  \country{New Zealand}}
\email{samuel.mann@op.ac.nz}

\author{Ruth Myers}
\affiliation{%
  \institution{Otago Polytechnic}
  \country{New Zealand}}
\email{ruth.myers@op.ac.nz}

\author{Dave Guruge}
\affiliation{%
  \institution{Otago Polytechnic}
  \country{New Zealand}}
\email{dguruge@me.com}

\author{Lucky Hawkins}
\affiliation{%
  \institution{Otago Polytechnic}
  \country{New Zealand}}
\email{lucky.hawkins2025@gmail.com}

\author{Kylie McKee}
\affiliation{%
  \institution{Dairy Australia}
  \country{Australia}}
\email{kylie.mckee@dairyaustralia.com.au}

\author{Rex Alexander}
\affiliation{%
  \institution{Envirocom}
  \country{New Zealand}}
\email{rex.alexander@envirocom.co.nz}

\author{Jamie Vaughan}
\affiliation{%
  \institution{Otago Polytechnic}
  \country{New Zealand}}
\email{pebkac@snap.net.nz}

\author{Tim Lynch}
\affiliation{%
  \institution{Otago Polytechnic}
  \country{New Zealand}}
\email{Timothy.Lynch@op.ac.nz}

\author{Danny Fridberg}
\affiliation{%
  \institution{Otago Polytechnic}
  \country{New Zealand}}
\email{Danny.Fridberg@op.ac.nz}

\renewcommand{\shortauthors}{Mann et al.}

\begin{abstract}
  Computing’s dominant concepts—innovation, efficiency, resilience, professionalism—often migrate from reflective ideals to instruments that limit behaviour, redirect responsibility, and foreclose critique. We call this drift weaponisation: the discursive repurposing of professional concepts so that they stabilise business‑as‑usual while making structural alternatives appear unreasonable, illegible, or out of scope. Using collective autoethnography across education, justice, public administration, research management, and computing, we identify recurring mechanisms (simplification, individualisation, binary framing, metric substitution, hero/resilience scripts, organised ignorance). From this synthesis we propose Reframing  — peaceful, practice‑ready shifts (e.g., From Simplified Slogans to Structural Literacy; From Performative Compliance to Meaningful Outcomes; From Coping to Justice)—each paired with a “do‑now” prompt. The levers restore complexity, surface assumptions, and redirect attention to structural conditions without requiring formal authority. We contribute: (1) a cross‑field account of weaponisation as a patterned phenomenon; (2) a portable reframing toolkit grounded in narrative and systems thinking; and (3) implications for computing within limits, including day‑to‑day practice shifts toward care, sufficiency, and justice.
  
\end{abstract}


\maketitle

\section{Introduction}

Christoph Becker’s \emph{Insolvent: How to Reorient Computing for Just Sustainability} \citep{becker2023insolvent} argues that meeting today’s ecological and social crises requires more than technical innovation: it demands a transformation in the professional practices, values, and narratives that organise computing work. Contemporary computing remains “insolvent,” Becker contends, because dominant norms continue to prioritise efficiency, growth, and optimisation over justice and sustainability; what is needed is a broader socio‑ecological orientation that integrates critical social theory, systems thinking, and care into practice. Yet efforts to reorient computing are often undermined by the very concepts meant to guide reform, which are themselves vulnerable to weaponisation.

By \emph{weaponisation}, we mean the process by which an idea is reframed and deployed to limit rather than to understand: nuance is stripped, complexity collapsed, and the concept becomes a simplified moral instrument used to redirect responsibility, constrain debate, or silence alternative perspectives. In doing so, weaponised narratives subtly narrow what appears legitimate or realistic, shrinking the space of possible responses available to professionals — even if the language of improvement or reform remains in place.  

To be clear, we are not arguing for weaponisation, nor even using weaponised in a positive sense. Our use of weaponised is in the sense of “weaponised against us” -- where "us" are communities \cite{Tupaea}, indigenous mothers \cite{Moyle25}, minority communities \cite{Isaac}, language revitalisation \cite{Jacob&Sabzalian}, victims \cite{Chouliaraki}. activists \cite{Daramy25} and so on.  Our argument is that we need to be able to recognise this framing, and to suggest approaches for reframing.  This paper explores how such weaponisation plays out across multiple fields of professional practice. We then adopt a stance borrowed from peace-building  of interrupting such narrative closure and replacing it with alternative narratives. 

Vodeb and Escobar \citep{Vodeb&Escobar}, describe an extra-disciplinary approach of extending beyond epistemological boundaries, learning from even "faraway'' disciplines and social movements, "what’s crucial is to go back. To go back to the discipline with the aim to change it".  Our approach, therefore, is to explore weaponisation in a variety of professional fields, to learn how interrupting narrative closure and alternative narratives function in practice, then to bring the lessons learned back as implications for the LIMITS community. 

We are mid-career professionals with experiences spanning education, justice, public administration, agriculture, research management, and computing, offering a cross‑sectoral view rarely captured in computing research. Each co‑author contributes a case from their domain, illustrating how a concept becomes weaponised and with what consequences for practice and understanding. Taken together, these cases suggest that weaponisation is not a local anomaly or sector‑specific quirk, but a patterned way of shaping professional imagination and responsibility across contexts.

To examine these patterns, we adopt collective autoethnography as a collaborative, reflective method grounded in lived professional experience. We treat our diverse practice worlds as analytically rich sites where narrative power, structural constraints, and institutional habits become visible. Through shared reflection and iterative dialogue, we generate a shared understanding of how weaponised narratives operate and how they might be disrupted.

In response, we introduce a set of \emph{reframing levers}: progressive, peaceful disruptions designed to unsettle weaponised narratives without escalating conflict or relying on heroic resistance. These devices restore complexity, surface hidden assumptions, and redirect attention toward structural conditions. By applying them across contexts and then returning the insights to computing, we examine how computing practice can better respond to the ecological and social limits highlighted in LIMITS scholarship.

The contribution of this paper is threefold. First, we identify a recurring pattern of narrative weaponisation across multiple fields of professional practice. Second, we offer a practical framework of reframing devices that practitioners can use to expand what is sayable, thinkable, and doable in their contexts. Third, by bringing cross‑field insights back to computing, we outline implications for computing practice and education within limits.

Recognising and disrupting weaponised narratives is an important step toward building computing professions capable of working within ecological and social limits. Without the ability to reframe harmful concepts, professionals risk remaining caught in coping rather than structural change, and narrow performance metrics rather than meaningful socioecological outcomes. 
We see this as precisely the terrain where LIMITS invites computing to imagine and enact different futures.

\section{Background and Related Work}
\label{sec:background}

\subsection{Narrative power and the weaponisation of concepts}
Across organisational studies and critical theory, narratives can be understood as organising structures that make some courses of action appear natural, necessary, or responsible. Sensemaking theory shows how organisational actors move from ambiguity to action through shared accounts of what is happening and what the situation demands \citep{weick2005organizing}. Organisational narrative research extends this insight by showing how stories stabilise identity, legitimate change, and define what can be said and done within organisational settings \citep{boje1991storytelling,vaara2016narratives}. Critical work on framing, myth, and technological ideology adds that such narratives also distribute responsibility by naming causes, agents, victims, and remedies, while leaving other relations unnamed \citep{benford2000framing,mosco2004digital,ames2019charisma}. This is the context in which concepts such as professionalism, efficiency, innovation, or resilience can become dangerous. When stripped of nuance and redeployed to restrict behaviour or deflect critique, they cease to function as prompts to reflection and become instruments of discursive control. We call this process \emph{weaponisation}.

The term \emph{weaponisation} originally referred to the process of converting an object or system into a weapon or using it as an instrument of force. In contemporary political and scholarly discourse, however, the term has broadened considerably, often describing the strategic use of symbolic, legal, or communicative resources as instruments of pressure or political advantage \citep{mattson2020weaponization,gelber2023weaponisation}. Our use of the term draws on traditions that analyse how power operates through discourse and symbols. Foucault argues that discourse does not merely describe reality but structures what can be thought, said, and done within a field of practice \citep{foucault1980powerknowledge}. Fairclough’s critical discourse analysis shows how everyday language can reproduce ideology and stabilise institutional power by making particular assumptions appear natural or commonsensical \citep{fairclough1992discourse}. Bourdieu similarly describes symbolic power as the capacity to impose meanings that appear legitimate and taken for granted \citep{bourdieu1991language}. In this paper we therefore use \emph{weaponisation} in a discursive sense: to describe the process by which professional concepts or values are stripped of nuance and redeployed as rhetorical instruments that discipline behaviour, redirect responsibility, and foreclose critique. In LIMITS terms, such narrative closure stabilises business-as-usual trajectories by making structural alternatives appear unreasonable, illegible, or out of scope. In public-facing policy and media, this reframing often individualises structural problems; Ennis documents how corporate ``dark PR'' recasts systemic harms as matters of personal choice (e.g., ``every little bit counts''), narrowing the range of legitimate responses and obscuring upstream levers \citep{ennis2023darkpr}.

Shotwell’s account of \emph{epistemic responsibility} in \emph{Fierce Love} adds to this critique by showing how AIDS activists developed practices that bridge lay and professional expertise and generate virtuous epistemic effects at the collective level—an antidote to purity logics that shame or silence \citep{shotwell2016fiercelove}. This communal stance underwrites our approach: responsibly ``knowing together'' counters the disciplining force of weaponised narratives by expanding who counts as a knower and what counts as admissible evidence \citep{shotwell2016fiercelove}.

\subsection{Professional practice under pressure}
Weaponisation flourishes under audit and performance cultures— KPI dashboards, compliance rituals, standardised metrics—that privilege speed, quantification, and surface appearance over structural inquiry. In such environments, slogans like ``efficiency,'' ``excellence,'' or even``resilience'' legitimise resource cuts, intensify workloads, or mask inequities while maintaining an aura of improvement \citep{nardi2018computing}. Ennis says these pressures translate into victim‑blaming frames that offload responsibility onto individuals and encourage ``ethical consumer'' detours instead of systemic remedies \citep{ennis2023darkpr}. Shotwell, drawing on Lorraine Code’s ecological thinking, reframes responsible knowing as embedded, communal, and historically attentive, which directly resists audit‑friendly simplifications \citep{shotwell2016fiercelove}.

\subsubsection*{Illustrative case: the weaponisation of ``resilience'' and ``self-care''}
Across health and care professions, ``resilience'' and ``self-care'' are often framed as \emph{individual duties} rather than system obligations, shifting attention away from staffing, workload, and organisational responsibility. A growing literature shows how this responsibilisation individualises structural harms and normalises coping over change. Critical reviews of COVID-era care work and burnout rehabilitation demonstrate how workers are positioned as self-responsible rehabilitees, while broader analyses frame resilience as a governmental technology that shifts risk from institutions to individuals and aligns with austerity and performance logics \citep{palecka2023whocares,korhonen2021individualizing,korhonen2021emc,welsh2014resilience,howell2015resilienceNJ}. Empirical studies further show that individual self-care is often constrained or impossible in practice, with workers over-functioning in ``morally uninhabitable'' environments, while ``hero'' narratives and ideals of self-management obscure organisational accountability \citep{lewis2022ssmmh,loyal2025socscimed,turcotte2024ninq,thorn2022selfmanaging}. Read through a multi-level lens, weaponised resilience operates across structural, institutional, symbolic, ideological, interpersonal, internalised, and embodied levels. The implication for LIMITS is clear: shifting from moralised individual coping to system accountability and collective capability aligns with just, sufficiency-oriented computing.

\subsubsection{Computing’s myths, debts, and \emph{problemism}}
Within computing, Becker argues that three interlocking myths: value‑neutral technology, rational decision‑making, and objective problems—under\-write a fourth myth of solvency (the belief that computational problem‑\-solving inherently improves the world) \citep{becker2023insolvent} (see also \citep{ToyamaHeresy, Mann23ComputingSoul}). Together, these sustain \emph{problemism}: a rationalist preoccupation with framing and solving ``problems'' that obscures how data and problem definitions are socially constructed and politically contested, thereby incurring ``debts of computing'' to societies and ecologies \citep{becker2023insolvent}. Becker proposes reorienting toward \emph{just sustainability design} by drawing on ``critical friends'' (critical social theory, feminist thought, systems thinking) and by asking explicitly which values should become facts in the systems we build. This repositioning complements our project: both seek to surface assumptions, reopen plural perspectives, and connect design choices to justice‑oriented outcomes.

\subsubsection{LIMITS scholarship and alternative imaginaries}
LIMITS scholarship interrogates narratives of innovation, disruption, scalability, optimisation, and neutrality, arguing that they align with extractive, high‑energy trajectories and advocating for computing grounded in care, sufficiency, and justice \citep{nardi2018computing}. It has elevated maintenance/repair and non‑linear design as antidotes to growth‑centric framings, while documenting how dominant stories crowd out viable alternatives. Our focus on weaponisation explains why such alternatives struggle to stick: familiar concepts are routinely recoded to stabilise the status quo \citep{nardi2018computing}. In parallel, the TechOtherwise Collective calls for moving beyond the binary of ``tech or not'' toward making tech differently in the service of collective wellbeing, an explicit invitation to re‑story practice \citep{techotherwise2021defund}. Degrowth literature similarly argues that to ecologise society we must imagine and enact alternatives to growth‑based development, not simply implement ``greener'' versions of the same paradigm \citep{kallis2015degrowthAlternative}.  One such imagining Escobar calls for, is "Pluriversalising technology" acknowledging "radical interdependence" \citep{Escobar2025}.

\subsubsection{Gaps addressed by this work}
\textbf{Cross‑field patterning.} Weaponisation is observed in many sectors, but rarely compared; we address this through collective autoethnography spanning education, justice, public administration, research management, and computing. 
\textbf{Actionable tools.} Diagnoses outnumber techniques; our reframing devices provide portable, do‑now prompts for practice aligned with LIMITS’ orientation \citep{nardi2018computing}. 
\textbf{Bridging to computing within limits.} We bring cross‑domain insights back to computing to support practice and pedagogy responsive to ecological and social limits \citep{nardi2018computing}.

\section{Method: Analytic Collective Autoethnography}\label{sec:method}

We approach this study as an analytic collective autoethnography. Autoethnography is not simply reflection on personal experience; it is the systematic analysis of lived experience as a route into understanding cultural, organisational, and professional worlds \citep{ellis2011autoethnography}. In this paper, the relevant ``culture'' is professional practice: the shared language, assumptions, routines, documents, obligations, and institutional pressures through which practitioners come to understand what counts as responsible, reasonable, or possible in their fields. The method is autoethnographic because the empirical material is drawn from the authors' own situated participation in these professional worlds; collective because analysis was produced through dialogue across multiple practice domains; and analytic because our aim is to develop a transferable account of a broader social phenomenon rather than to present personal reflection as an end in itself. Following Anderson's account of analytic autoethnography, we position the authors as complete member researchers: practitioners with sustained participation in the fields being examined, visible presence in the analysis, reflexive engagement with their own locations, and a commitment to theoretical development beyond the individual case \citep{anderson2006analytic}. This stance also aligns with insider and work-based research, where practitioners investigate questions arising from their own fields of practice while treating their professional knowledge, documents, relationships, and routines as legitimate sources of inquiry \citep{costley2010doing}. It further aligns with collective autoethnography as a method for generating insight through the comparison and interrogation of situated experience across multiple researchers \citep{chang2013collaborative}.

Each author contributed from a position of "necessarily insider" \cite{mannEthicalBecoming25}  rather than external observation. Several authors are also drawing on larger autoethnographic studies of professional practice within their own fields of practice. These studies examine how practitioners make sense of their work, negotiate institutional expectations, and develop professional judgement and ontological becoming over time \cite{MannBecoming26}. The material brought into this paper therefore sits within a wider body of sustained reflective inquiry, rather than being generated only for this article.

The reflections draw on long-term participation in the authors' respective fields and on the kinds of records through which professional practice is ordinarily remembered, contested, and interpreted: reflective notes, teaching and assessment materials, policy and strategy documents, professional frameworks, meeting notes, project documentation, public organisational documents, and repeated experience of similar professional situations. In this sense, the material analysed here should be read neither as detached observation nor as isolated anecdote \citep{schon1983reflective,anderson2006analytic,costley2010doing}. The cases are analytically selected incidents and recurring professional episodes that condense patterns encountered over time within particular fields of practice. 

Data generation and analysis proceeded in three stages. First, each author identified one to three incidents (which may be slow-burn) or recurring episodes in which a valued professional concept, such as professionalism, fairness, sustainability, resilience, compliance, efficiency, or education, appeared to narrow what could be said, done, or imagined. Authors were asked to select material that was specific enough to show how narrative closure occurred in practice, while also sufficiently anonymised to protect individuals and organisations. Second, authors prepared short reflective accounts using a shared template: professional setting; concept or value invoked; incident or recurring practice in which the concept was mobilised; what became difficult to say or do; where responsibility was located or displaced; consequences for practice; and possible reframing. This template was used to retain incident-level texture while making the accounts comparable across fields.

Third, these accounts were shared among the author group for peer annotation, discussion, and cross-case analysis. Co-authors read each other's accounts as practitioner-analysts bringing situated expertise to the material. Annotations focused on the mechanism of closure, the level at which responsibility was being shifted, the role of institutional language or procedure, and the questions that might reopen the situation. This stage was important because several contributors initially described their cases in the ordinary language of their field. Group discussion helped make visible the narratives, assumptions, and mechanisms that had become normalised within that field.

Through iterative discussion, we compared the incidents across domains and grouped the recurring mechanisms into a provisional analytic vocabulary: simplification, individualisation, binary framing, metric substitution, symbolic substitution, procedural drift, organised ignorance, hero or resilience scripts, and displacement of responsibility. These categories were refined by returning to the incident accounts and asking whether the category helped explain what had become unsayable, unthinkable, or out of scope in each case. The analysis therefore moved between incident, author interpretation, peer response, cross-case comparison, and theoretical synthesis. This process produced the account of the patterns of weaponisation developed in Section 5.

Once we had developed this shared understanding of how weaponisation could frame and explain the patterns across the cases, we developed the disruptive reframing prompts presented in Section 6.1. These prompts emerged from two sources. Some were drawn from moves that contributors were already using in practice to reopen situations, surface assumptions, or shift attention back toward structural conditions. Others were developed through the reframe-and-replace approach introduced in Section 6: rather than simply countering a weaponised narrative, we asked what alternative framing could displace it with a more generous, structurally aware, and practically usable story. For each mechanism of weaponisation, we asked what small, peaceful, practice-ready move could interrupt narrative closure without requiring formal authority or escalating conflict.

The resulting levers were then tested back against the incident accounts. Each lever had to meet at least one of three criteria: make visible a hidden assumption, redirect attention toward structural conditions, or open a more generous account of professional responsibility. The levers are therefore not presented as analytic products of the collective autoethnographic process and as prompts for further use, critique, and refinement.

This method shapes how the findings should be read. First, the cases are analytically selected incidents from experienced practitioners positioned within those sectors - they are not necessarily representative of those sectors. Second, the claims are not about the prevalence of weaponisation across all organisations; they concern the mechanisms by which weaponisation becomes recognisable across different fields of practice. Third, the value of the method lies in the relation between situated texture and cross-case patterning: the incidents show how narrative closure takes place, while the synthesis identifies mechanisms that may be useful for computing practitioners working within ecological and social limits.

Given the qualitative design, we follow Tracy's qualitative quality criteria by foregrounding sincerity, credibility, resonance, and usefulness \citep{tracy2010,Lopeztracy2020}. Sincerity is addressed through author visibility and explicit acknowledgement of insider positioning. Credibility is addressed through a shared incident template, peer annotation, traceable movement from incident to mechanism, and collective interrogation of emerging categories. Resonance is sought through thick enough incident accounts to make the texture of narrative closure recognisable to practitioners in other settings. Usefulness is pursued through the development of reframing levers that translate analysis into practice-oriented prompts. Our aim is to provide tools that practitioners can wield immediately to disrupt weaponised narratives peacefully, while inviting further empirical work to evaluate effects in specific settings.

\subsubsection{Ethics, scope, and limitations}

Ethically, the study did not involve recruiting external participants or collecting data about identifiable third parties. The authors wrote from their own professional experience and removed or generalised details that could identify individuals, organisations, or sensitive situations. Where professional documents or frameworks are discussed, they are treated as contextual artefacts through which practice is interpreted, rather than as confidential data sources.  Limitations include cultural situatedness, potential selection bias in cases, and the generative (rather than prescriptive) nature of the framework. We mitigate these by (i) inviting plurality across fields, (ii) making our reframing prompts explicit for critique and reuse, and (iii) returning implications to computing where they can be tested in curricula and practice \citep{nardi2018computing}.
\section{Contexts of Weaponised Concepts}
\label{sec:contexts}

This section presents a series of short reflective contributions from co-authors, each drawn from their own domain of professional practice. For reasons of space, the accounts are presented in condensed form. This allows the shared patterns across cases to come into focus, while inevitably reducing some of the stylistic variation, situated detail, and multi-vocal texture of the original contributions. The aim is not to provide exhaustive empirical accounts, but to surface recurring patterns through which professional concepts become weaponised. These reflections show how ideas that originally function as guides for professional judgement — such as sustainability, biosecurity, legislative protection, fairness, professionalism, and education — can be reframed and redeployed to limit behaviour, redirect responsibility, or narrow what responses appear legitimate.

Each contribution identifies a concept or value that plays an important role in its field and describes how it becomes mobilised in practice. In each context, weaponisation occurs through processes such as simplification, binary framing, symbolic substitution, administrative drift, or institutional pressure. The reflections also attend to what becomes obscured when this happens, including structural conditions, invisible labour, professional identity, and unequal distributions of risk, responsibility, and harm.

These contributions serve two purposes. First, they demonstrate the cross-domain character of weaponised concepts, showing that similar dynamics recur across otherwise very different professional contexts. Second, they introduce the conceptual patterns developed in the following section, where these dynamics are brought together and analysed more explicitly.

\subsubsection{Biosecurity and Responsibility in Rural Systems}

Within biosecurity and One Health systems \citep{Windker2025_Onehealth}, responsibility for complex ecological and health risks is often translated into simplified narratives of individual behaviour and compliance.

Biosecurity and One Health frameworks seek to address complex interactions between human health, animal health, and environmental systems. These frameworks acknowledge that disease emergence, ecological disruption, and food system risks are shaped by interconnected global processes involving agriculture, trade, land use, and environmental change.

In practice, however, responses to biosecurity threats often become simplified into narratives focused on individual behaviour. Responsibility for preventing disease spread or ecological disruption can be framed primarily in terms of compliance, awareness, or individual decision-making. These narratives emphasise what individuals or communities should do differently while the broader structural dynamics shaping risk remain less visible.

When systemic challenges are translated into behavioural narratives, responsibility is subtly displaced downward. Structural drivers such as global supply chains, regulatory arrangements, and institutional decision-making recede from view, while individuals become the primary site of intervention. This simplification can obscure the complexity of the systems involved and narrow the range of responses that are considered possible.

Within this dynamic, biosecurity discourse may unintentionally reinforce a flattened understanding of responsibility, where complex ecological systems are reduced to matters of compliance or personal accountability.  But we must also recognise a `bi-directional weaponisation' - sometimes both parties are just not willing to let go of whatever barriers, blockages or stuckness they face, and we see what I call `stall loops' come in where the “blame game” shifts, and the ownership of the narrative is claimed by one party, and the opportunity for progressive navigation ends. 

\subsubsection{Sustainability and Invisible Labour}

In culinary practice, responses to climate change increasingly depend on forms of labour that remain largely unrecognised within institutional structures. Activities such as maintaining relationships with ethical suppliers, coordinating community engagement, managing waste streams, mentoring junior staff, and sustaining collaborative learning environments are essential to the functioning of sustainability initiatives. Yet this relational and emotional work is rarely captured within conventional measures of productivity or formal workload allocations.

At the same time, institutions derive increasing symbolic value from sustainability commitments. Initiatives such as carbon-neutral campuses, zero-waste programmes, and regenerative food practices contribute to organisational legitimacy and public reputation. Within this environment, the labour required to sustain these initiatives can become obscured. Sustainability narratives remain highly visible, while the everyday work required to enact them remains largely hidden.

This creates a dynamic in which sustainability labour is quietly absorbed by individuals rather than supported by institutional structures. Activities necessary to sustain ethical food systems are reframed as expressions of passion, commitment, or professional identity rather than recognised as work requiring time, resources, and organisational support. In this way, the relational and emotional labour underpinning sustainability initiatives can become a silent subsidy to the institution, generating symbolic value while the practical costs remain unevenly distributed.

\subsubsection{Legislative Protection and Volunteer Firefighters}

New Zealand’s Accident Compensation scheme was originally designed as a universal system of protection, yet its contemporary interpretation reveals tensions between legislative ideals and the lived realities of volunteer emergency service workers.

New Zealand’s Accident Compensation system was designed as a universal, no-fault approach to injury support and rehabilitation. The scheme removed the right to sue while providing broad coverage for physical injury across work, sport, and everyday life. In principle, it represents a collective commitment to shared protection and social responsibility.

However, significant gaps remain when the legislation is applied to work-related illness and psychological harm. Volunteer firefighters, who make up approximately 86 percent of New Zealand’s firefighting workforce, do not receive the same coverage as their professional counterparts for psychological trauma or occupational disease arising from emergency service work. Although these volunteers operate under the same hazardous conditions, the legal framework defining employment relationships excludes them from key protections.

The resulting situation highlights a tension between the universal ideals of the scheme and its administrative interpretation. Legislative and procedural reasoning emphasises fiscal sustainability, definitional boundaries, and legal precedent. Yet these interpretations sit uneasily alongside the scale of civic contribution provided by volunteer emergency services. The system that was originally conceived as a universal safeguard can, in practice, produce uneven protection across those performing similar work under similar risks.

\subsubsection{Fairness and Organisational Process}

Formal complaint and disciplinary procedures are intended to safeguard fairness in professional organisations, yet the authority of process itself can sometimes reshape organisational dynamics in unexpected ways.

Formal procedures play an important role in organisational life. Complaint processes, disciplinary procedures, and governance frameworks are intended to provide safeguards against arbitrary decision-making and to ensure fairness in professional environments.

In practice, however, these same procedures can sometimes be mobilised in ways that shift them from mechanisms of fairness into instruments of organisational discipline. Formal complaint processes may be initiated following relatively minor disagreements, triggering investigative procedures, documentation requirements, and reputational pressures that extend far beyond the substance of the original issue. Once activated, the administrative machinery of the process can reshape organisational dynamics regardless of the eventual outcome.

Because organisations often emphasise procedural neutrality, leadership responses frequently focus on allowing “the process to take its course.” While this stance reinforces the legitimacy of formal rules, it can also obscure how procedures are functioning in practice. The authority of the process itself becomes the primary organising force, and the possibility that procedures may be misused strategically can remain difficult to discuss openly.

In this way, systems designed to protect fairness may inadvertently become mechanisms through which pressure is applied, reputations are shaped, and organisational power relations are negotiated.

\subsubsection{Professional Identity in the IT Profession}

Within the technology profession, widely repeated explanations about service roles, skills shortages, and communication challenges can gradually solidify into narratives that shape how professional expertise is recognised and exercised.  Several familiar explanations are often used to describe tensions between IT teams and the rest of the organisation. These explanations typically appear as common-sense observations: that IT is a service function, that the industry suffers from a shortage of skills, or that communication between technologists and leadership needs improvement.

Each of these ideas contains elements of truth. However, over time they can become simplified into slogans that quietly define the boundaries of the profession. When IT is consistently framed as a service function, the organisation becomes the customer and technology professionals become providers of a utility. Professional judgement, architectural concerns, and long-term system risks are then positioned as issues of customer service rather than matters of professional expertise.

Similarly, the persistent narrative of a skills shortage can shift attention toward individual capability development while obscuring organisational structures that limit the influence of technical expertise in decision-making. Communication narratives can reduce complex structural relationships between technology, governance, and organisational strategy to matters of interpersonal skill.

Across these narratives, useful explanations gradually become disciplinary scripts. Ideas that once helped people understand the profession begin to shape and constrain it, subtly influencing whose expertise is heard and whose role is treated as optional.

\subsubsection{Education and the Logic of Control}

In many contemporary educational environments, systems designed to measure and optimise learning increasingly shape how students and teachers understand engagement, responsibility, and success.

Education is often presented as a pathway toward growth, curiosity, and personal transformation. Yet within many institutional environments the structures surrounding learning emphasise measurement, surveillance, and compliance. Attendance tracking, behaviour monitoring, performance metrics, and data-driven decision systems are introduced as tools intended to improve learning outcomes and institutional accountability.

When these mechanisms become dominant, however, learning environments can begin to resemble systems of control rather than spaces of exploration. Success becomes closely tied to targets, scores, and institutional indicators, while curiosity, experimentation, and relational trust receive less attention. The language used to explain student behaviour can reinforce this shift. When learners disengage, the explanation often defaults to a simple narrative: they are not interested in learning.

Such explanations obscure the broader conditions shaping learner responses. Mandatory attendance, previous educational harm, and the pressures of institutional systems can create environments where withdrawal becomes a rational response. In these contexts, educational structures designed to support learning may instead contribute to experiences of surveillance, judgement, and exclusion.

As institutional systems prioritise optimisation, prediction, and measurable outcomes, the complex human realities of learning risk being flattened into numbers and compliance indicators.

\subsection{Relation to prior accounts of these dynamics}

The patterns identified here are not presented as wholly novel dynamics. Many have been documented within specific fields, often under different conceptual names. In agricultural biosecurity, for example, critiques of individualised approaches to farmer behaviour show how disease prevention can be framed through compliance and behavioural responsibility while overlooking cultural identity, care, and the practical conditions of farming work \citep{maye2020farm}. In food work, accounts of neoliberal labour conditions and culinary placemaking similarly show how relational, affective, and place-making labour can become precarious or under-recognised \citep{chartrand2025foodwork}. In education, scholarship on datafication and surveillance has shown how learning, childhood, and institutional accountability are increasingly mediated through data systems that render complex educational relationships into measurable indicators \citep{lupton2017datafied,jarke2019datafication,szcyrek2022surveillance}.

Our contribution is therefore not the claim that individualisation, datafication, invisible labour, or displaced responsibility have gone unnoticed. Rather, we bring these dynamics together across multiple professional fields and analyse them as related forms of narrative weaponisation. This framing shows how valued concepts such as biosecurity, sustainability, fairness, professionalism, education, efficiency, and resilience can be redeployed to close down inquiry, redirect responsibility, and stabilise existing institutional arrangements. The cross-field comparison makes visible a shared mechanism: concepts that appear constructive within a field can become restrictive when they are simplified, abstracted from their conditions, and used to narrow what professionals can say, question, or imagine. This is the basis for the reframing levers developed in Section 6.1, which are intended to move from diagnosis toward practice-ready forms of interruption and replacement.
\section{Patterns of Weaponisation}
\label{sec:patterns}

The contributions in Section~\ref{sec:contexts} span diverse domains of professional practice. Despite contextual differences, the reflections reveal several recurring mechanisms through which professional concepts become weaponised in practice. While the specific forms vary across contexts, similar dynamics appear repeatedly: responsibility is shifted downward, symbolic narratives substitute for material support, procedural systems drift from protection toward control, and professional identities become narrowed through simplifying scripts.

Across several contexts, complex systemic challenges are reframed as matters of individual behaviour or responsibility. In the biosecurity reflection, ecological and epidemiological risks emerging from global agricultural systems and environmental change are translated into narratives centred on compliance and personal decision-making. Similarly, in educational settings, disengagement is frequently interpreted as a lack of student motivation rather than a response to institutional structures or prior experiences of learning. Such narratives simplify complex systems into manageable behavioural interventions. While these framings can provide clear points of action, they also displace attention from structural conditions that shape outcomes. Responsibility becomes concentrated at the level of the individual while institutional arrangements, economic pressures, and governance systems recede from view.

A second recurring pattern concerns the relationship between symbolic commitments and the labour required to sustain them. In the sustainability reflection from culinary practice, organisations derive legitimacy from visible commitments to ethical food systems, waste reduction, and environmental responsibility. Yet the relational and organisational labour required to sustain these initiatives frequently remains unrecognised within formal workload structures. Activities essential to sustaining these initiatives may be reframed as expressions of professional passion or personal commitment rather than recognised as work requiring institutional support. In this way, organisational legitimacy can be maintained while the material conditions required to sustain ethical practice remain unevenly distributed.

Several reflections also reveal how procedural systems designed to protect fairness or provide support can gradually shift toward more defensive or disciplinary functions. The account of legislative protection within the New Zealand Accident Compensation system illustrates how administrative interpretations and legal reasoning can narrow the reach of frameworks originally intended to provide universal coverage. Similarly, organisational complaint and disciplinary procedures may operate in ways that extend beyond their intended protective role. Once formal processes are activated, the administrative machinery of investigation, documentation, and review can reshape organisational relationships regardless of the eventual findings. In such situations, the authority of process itself becomes a central organising force, sometimes obscuring questions about how procedures are functioning in practice.

A further pattern concerns the role of widely repeated professional narratives. In the information technology reflection, familiar explanations such as the framing of IT as a service function, the persistent narrative of skills shortages, or the emphasis on communication gaps between technologists and leadership all contain elements of truth. Over time, however, these explanations can become simplified scripts that define the boundaries of professional action. Professional judgement may be reframed as a matter of customer service, systemic organisational tensions reduced to communication issues, and structural constraints on professional agency obscured by calls for individual capability development.

These patterns suggest that weaponisation does not necessarily arise from deliberate manipulation or malicious intent. More often it emerges through ordinary organisational processes: simplification, institutional pressure, administrative interpretation, and the repeated circulation of apparently reasonable narratives. Concepts that initially function as guides for professional judgement gradually become instruments through which behaviour is shaped, responsibility is redistributed, and the boundaries of legitimate response are narrowed.

Across the reflections, a deeper dynamic is also visible. Many of the narratives operate within an extractivist logic in which institutions derive value from professional labour, civic participation, or environmental commitment while distancing decision-making structures from the social and ecological responsibilities associated with those activities. This structural separation is also visible in accounts of ICT and AI work: Widder and Nafus show how developers may recognise harms while locating responsibility elsewhere in a modularised “supply chain”, while Brooks et al. suggest that ICT professionals’ role in sustaining environmentally damaging industries has remained under-examined, requiring more direct professional reflection on when such systems should no longer be designed or maintained \citep{Widder,BrooksTech4Bad}. In this sense, weaponised narratives can function as mechanisms through which responsibility is rhetorically acknowledged while being practically displaced.

These dynamics are often sustained through forms of dissociation. Narratives of efficiency, professionalism, fairness, or individual responsibility allow actors within a system to separate everyday decision-making from its broader social consequences. Through this process, labour, risks, and harms can be externalised or rendered invisible. Social and ecological responsibilities may remain present at the level of language or aspiration while becoming detached from operational practice.

Such mechanisms also rely on processes of abstraction and othering. Complex social relationships are translated into categories, metrics, procedures, or behavioural expectations that make systems easier to administer but harder to question. The resulting structures can simultaneously position individuals as both subjects and agents of the system: affected by its constraints while also participating in its reproduction through everyday professional practice.

Recognising this dual position is important. Professionals often experience these narratives as pressures imposed upon them by institutional systems, yet they may also inadvertently reinforce them through the practical necessity of working within existing structures. In this sense, the dynamics described here do not divide neatly into victims and perpetrators. Rather, they reflect systems in which individuals are frequently both.

Understanding weaponisation in this way shifts the focus from identifying blame toward examining how professional narratives operate within organisational environments. The following section therefore turns to the question of reframing: how such narratives might be disrupted or reopened in ways that expand professional agency rather than narrow it.

\section{Reframing Weaponised Concepts}
\label{sec:reframing}

The reframing levers developed in this project are one response to that question. They are designed as practical ways of interrupting weaponised narratives without simply replacing them with counter-slogans or escalating conflict. Instead, they aim to reopen what weaponisation tends to close down: complexity, structural visibility, professional judgement, and the possibility of alternative futures.

Framing is not simply a matter of presentation or rhetoric. Across organisational and management research, frames shape how situations are interpreted, what responsibilities are recognised, and which actions appear legitimate or necessary \citep{cornelissen2014framing}. In complex public and social problems, reframing is especially important because problem and solution spaces co-evolve in non-linear ways, requiring systemic and iterative shifts in understanding rather than simple problem definition \citep{vanderbijlbrouwer2019problem}. Framing is also political: frames privilege some meanings, values, and responses while marginalising others, thereby helping to stabilise dominant norms and institutional arrangements \citep{prendeville2022politics}. Policy work further suggests that frames shape not only how present problems are understood, but also how actors imagine desirable, feared, or possible futures \citep{miedzinski2018policy}.

Peacebuilding and conflict transformation scholarship strengthens this argument by showing that harmful social dynamics are sustained not only through material structures but also through the narratives that justify, normalise, and reproduce them. In intractable conflicts, dominant conflict-supporting narratives help maintain antagonism, victimhood, and moral certainty, while movement towards peace requires the emergence of more plural and peace-supporting narratives \citep{bartal2014narratives}. While professional practice is not war, many sectors sustain conflict-supporting professional narratives—e.g., efficiency over care, innovation over maintenance—that normalise harm until they are pluralised and reauthored. Conflict transformation therefore depends not only on resolving immediate disputes but on changing the stories through which people understand themselves, others, and the conditions of possible action \citep{lederach2005moral,federman2016narrative}. Storytelling and reframing are especially important here because narratives can either intensify division and mask injustice or foster mutual recognition, empathy, and more constructive forms of social engagement \citep{senehi2002constructive}.

Beyond “countering,” practitioners in addressing violent extremism also emphasise replacement: ''counter‑narratives have limited utility and violent narratives must ultimately be replaced, not just countered'' \citep{BeutelNarratives}. The implication for professional practice is to prioritise coherent alternative narratives - stories that are emotionally resonant, structurally aware, and usable - so they can displace weaponised concepts rather than merely debate them \citep{BeutelNarratives}.

Recent scholarship on everyday resistance extends this picture by showing that politically meaningful change does not always depend on dramatic confrontation or heroic acts \citep{girei2023everyday}. Small-scale and often quiet forms of micro-resistance can unsettle dominant managerial and institutional logics, especially when they accumulate over time or connect to wider transformative agendas . Work on radical incrementalism similarly argues that modest corrective actions can contribute to broader structural change from within organisations and professional systems \citep{timorshlevin2023radicalincrementalism}. Narrative and discursive interventions are an important part of this process: reframing, deconstruction, and the subtle reshaping of stories can alter the discursive landscape in ways that make alternative futures more thinkable and actionable \citep{riedy2022discursive}. Accordingly, our \emph{Reframing Levers} are designed as micro-interventions that practitioners can enact now, accumulating into forms of radical incrementalism over time \citep{timorshlevin2023radicalincrementalism}. They serve as peaceful, progressive disruptions:  invitational moves that widen perception, unsettle harmful defaults, and open pathways for more just, structurally aware action.

A critical thread in systems thinking insists that understanding begins by seeing through others’ eyes, and continues by recognising the limits of our own worldview—Churchman’s provocation about the systems approach \citep{churchman1979systems}. We operationalise this via a multi‑level lens: \textbf{structural} (law, policy, funding), \textbf{institutional} (processes, norms), \textbf{symbolic} (categories, stories, defaults), \textbf{ideological} (assumptions of neutrality/merit), \textbf{interpersonal} (voice, recognition), \textbf{internalised} (self‑limiting scripts), and \textbf{embodied} (stress, silence, exclusion) \citep{mannKaretai_decolonialStory}. Histories of AIDS advocacy illustrate how shifting epistemic conditions—who is recognised as a knower and how problems are framed—can change material outcomes, underscoring why single‑level fixes rarely suffice \citep{shotwell2016fiercelove}. Recent design and sustainability research similarly links narrative change to deeper leverage points for eco‑social transformation \citep{vervoort2024nineDimensions}.

\subsection{Disruptive reframing prompts}
These reframing prompts emerged from reflections in Sections~\ref{sec:contexts} and~\ref{sec:patterns}  and iteratively developed in reflective application in practice (Section~\ref{sec:reframing_practice}).  For clarity, they are presented first here. 

\subsubsection{From Simplified Slogans to Structural Literacy}

Weaponised narratives thrive on simplicity; systems understanding dissolves their power.

\textbf{Action prompt:} Take a recurring problem. Map three upstream causes you normally ignore.

\textbf{Why this disrupts:} Because complexity exposes the mechanisms weaponised narratives work hard to hide.

\subsubsection{From Reactive Outrage to Curious Inquiry}

Weaponisation feeds on emotional escalation; curiosity disrupts the script.

\textbf{Action prompt:} Write down a recent moment of disagreement and ask yourself: ``What curiosity could I bring to this instead of certainty?''

\textbf{Why this disrupts:} Because curiosity shifts the interaction from conflict to understanding, breaking the cycle weaponisation relies on.

\subsubsection{From Performative Compliance to Meaningful Outcomes} 
Ritual, bureaucracy, or ``tick-box'' behaviours are often enforced or encouraged by weaponised ideas. 

\textbf{Action prompt:} Pick a KPI or metric you work with. What system-level outcome would be a truer measure of success?

\textbf{Why this disrupts:} Because focusing on real impact undermines the ritualised behaviours that keep harmful systems in place.

\subsubsection{From Hidden Frames to Visible Assumptions}

Weaponisation smuggles norms in as ``common sense''.

\textbf{Action prompt:} Choose one commonly used concept in your field and list the assumptions embedded inside it.

\textbf{Why this disrupts:} Because naming assumptions reveals the power dynamics baked into ``common sense''.

\subsubsection{From Individual Blame to System Insight}

Many organisational narratives place responsibility on individuals even when systems create the conditions.

\textbf{Action prompt:} Rewrite a blame-heavy explanation in system terms: incentives, constraints, histories, power.

\textbf{Why this disrupts:} Because reframing problems as systemic removes the moral force of blame and redirects attention to structural change.

\subsubsection{From Binaries to Complexity}

Weaponised narratives rely on false oppositions (strategic/operational, theory/practice, innovation/compliance).

\textbf{Action prompt:} Take one common binary in your profession and redraw it as a continuum, ecosystem, or cycle.

\textbf{Why this disrupts:} Because dissolving false oppositions prevents narrow narratives from controlling what counts as legitimate.

\subsubsection{From Flat Narratives to Multi-Level Understanding}

Weaponisation often collapses problems into a single level (usually interpersonal or individual), obscuring how forces operate across layers.

\textbf{Action prompt:} Pick one harmful narrative in your field. Identify one action you could take at each system level, structural, institutional, symbolic, ideological, interpersonal, internalised, embodied, to disrupt it.


\textbf{Why this disrupts:} Because seeing how power operates across levels exposes where interventions are possible, and where responsibility truly sits.

\subsubsection{From Organised Ignorance to Shared Understanding}

Weaponisation depends on selective blindness, what gets excluded, minimised, or deemed ``not our business''.

\textbf{Action prompt:} Choose one ``off-limits'' topic in your field and write down a single honest question about it, then share that question with one colleague today.

\textbf{Why this disrupts:} Because gently surfacing the unspoken widens what is discussable, weakening the boundaries that protect harmful narratives.

\subsubsection{From Coping to Justice}

When people are forced to cope with broken systems, justice begins by removing the conditions that make coping necessary.

\textbf{Action prompt:} Spot a workaround someone uses to ``cope''. What would it take to remove the need for it entirely?

\textbf{Why this disrupts:} Because shifting from endurance to root-cause change challenges systems that depend on people quietly absorbing harm.

\subsubsection{From Fear-Driven Constraint to Courageous Imagination}

Weaponised narratives shrink horizons (``be realistic'', ``that's impossible''). Imagination is a quiet form of resistance.

\textbf{Action prompt:} Sketch a future version of your work or organisation that feels impossible, then circle one thing you could prototype tomorrow.

\textbf{Why this disrupts:} Because imagining alternatives breaks the illusion that the current system is inevitable.

\subsubsection{Cultivate a More Generous Story}

Take a narrative that currently induces shame, fear, or restriction. Rewrite it as a story about capability, solidarity, possibility, or transformation.

\textbf{Action prompt:} Take one limiting story in your field and rewrite it as a story about collective capability, dignity, or possibility, then share that reframing with someone today.

\textbf{Why this disrupts:} It replaces the disciplining function of the weaponised narrative with an invitational one.

\section{Positive Reframing in Practice}
\label{sec:reframing_practice}

The previous sections described how professional concepts can become
weaponised through simplification, binary framing, and institutional
pressures that redirect responsibility and narrow what responses
appear legitimate. We also introduced a set of reframing prompts
designed to interrupt these dynamics. All these interventions are currently underway in practice and the subject of ongoing professional practice research. 

Each contribution revisits one of the professional contexts described
earlier and applies a specific reframing prompt to the situation.
Rather than reproducing the full narratives, the examples focus on
how the prompt changes the interpretation of the problem and what
becomes visible once that shift occurs. Each example therefore
illustrates a reframing move in action, followed by a short reflection
from the author about how the reframing is functioning within their
professional context.

These examples are intentionally concise. Their purpose is not to
present complete interventions but to demonstrate how small shifts in
interpretation can interrupt weaponised narratives and reopen
possibilities for professional judgement and collective action.

\subsection{From Simplified Slogans to Structural Literacy}

In agricultural biosecurity and disease management, problems are often framed through simplified behavioural narratives such as “farmers
should just vaccinate” or “compliance failures cause outbreaks.”
These narratives position disease management primarily as an issue of individual responsibility while obscuring the wider ecological, economic, and institutional systems shaping decision making on farms.

Reframing the issue through structural literacy shifts attention away from individual behaviour and toward the system that produces disease risk. In this work, journey maps and One Health systems diagrams were developed to visualise how disease dynamics emerge through interactions among climate variability, wildlife movement, farm infrastructure, supply chains, labour availability, and regulatory settings. These artefacts made visible how vaccination decisions were embedded within wider structural conditions rather than isolated acts of compliance.

Once these relationships were mapped, the central question changed. Rather than asking “why did the farmer not vaccinate?”, the discussion shifted toward identifying where the system itself was creating barriers to effective prevention. Infrastructure gaps, coordination failures, and conflicting policy signals—previously hidden within behavioural explanations—became visible points of intervention.

\subsection{From Reactive Outrage to Curious Inquiry}

Institutional responses to professional failure frequently begin with
questions of fault and compliance. Investigations may start from
assumptions that someone must be responsible for the breakdown, which
can lead participants to defend their actions rather than explore the
conditions that shaped them.

A reframing prompt used in several contexts involved shifting these
conversations toward inquiry. Instead of beginning with questions such
as “why did this fail?” or “who was responsible?”, practitioners were
encouraged to begin with exploratory questions such as: “What is making this difficult in practice right now?”, and “What conditions shaped the decision that was taken?''

These questions redirected attention away from blame and toward the
situational conditions shaping professional judgement. Resource
constraints, institutional pressures, and coordination challenges
became discussable in ways that compliance-focused conversations had
previously suppressed. Accountability remained present, but the
conversation expanded to include the system within which the decision
had occurred.


\subsection{From Hidden Frames to Visible Assumptions}

Professional narratives often operate through assumptions that remain
implicit. In educational and organisational contexts, one such
assumption is that institutions provide expertise while communities
or learners receive it. This framing positions practitioners as
service providers and others as passive recipients.

Making these assumptions visible became a reframing move in its own
right. In this work, discussions were structured to explicitly
surface the assumptions embedded in service narratives. Once
articulated, these assumptions could be examined rather than quietly
reproduced.

One reframing involved introducing the metaphor of a shared waka to
describe professional practice. Rather than positioning practitioners
and communities in opposing roles of provider and recipient, the
metaphor emphasised that participants were navigating the same
system together, albeit from different positions of responsibility
and knowledge.


\subsection{From Individual Blame to System Insight}

Weaponised narratives frequently attribute complex failures to
individual shortcomings. In educational contexts, disengagement may
be interpreted as a lack of motivation or willingness to learn. In
governance contexts, complaint procedures may isolate individual
actors while leaving institutional conditions unexamined.

Applying the reframing prompt of system insight shifts attention from
individual behaviour to the structures shaping it. Rather than asking
why individuals failed, practitioners examine how policies, resource
constraints, disciplinary norms, and historical experiences shape the
choices available to them.

Within this reframing, behaviours previously interpreted as
resistance or apathy can be understood as responses to institutional
arrangements that do not fully accommodate the realities of those
involved. Responsibility is therefore redistributed across the
system rather than concentrated exclusively on individuals.


\subsection{From Flat Narratives to Multi-Level Understanding}

Professional problems are often described through single-level
explanations that ignore interactions across ecological,
institutional, and social systems. Such flat narratives simplify
complex realities but also limit the range of possible responses.

Reframing through multi-level understanding involves mapping how
issues operate simultaneously across several interacting levels—for
example, environmental conditions, regulatory frameworks,
institutional practices, and local decision making. In the work
presented here, system maps and journey diagrams were used to make
these layers visible and to show how interventions at one level can
produce consequences elsewhere in the system.

By recognising these interactions, practitioners were able to move
beyond linear explanations and consider coordinated responses across
multiple levels of the system.


\subsection{From Organised Ignorance to Shared Understanding}

In some professional environments, certain questions become difficult
or impossible to ask publicly. Sensitive issues may remain unspoken
because discussing them appears professionally risky or institutionally
inconvenient.

One reframing move involved explicitly posing what the author termed
the “off-limits question.” Rather than working around the issue, the
question was stated directly:

\begin{quote}
“If we know that firefighters experience psychological harm, what
institutional arrangements prevent them from receiving timely
support?”
\end{quote}

By naming the question openly, the reframing disrupted a pattern of
organised ignorance in which the issue was widely recognised but
rarely discussed directly.


\subsection{From Fear-Driven Constraint to Courageous Imagination}

Professional institutions often operate under narratives of
constraint—legal liability, procedural boundaries, or funding
limitations—that shape what practitioners believe is possible.

Reframing through courageous imagination asks what new arrangements
might become conceivable if these constraints are treated as design
problems rather than fixed limits. In the context explored here,
this reframing opened discussion about alternative medico-legal
support pathways capable of providing faster psychological support
for frontline firefighters while navigating existing regulatory
structures.


\subsection{Cultivating a More Generous Story}

Weaponised narratives frequently position participants in adversarial
roles—provider versus client, regulator versus regulated, expert
versus public. These narratives narrow the possibility of cooperation
by framing interactions primarily through compliance or conflict.

Reframing toward a more generous story expands the narrative frame
to recognise shared challenges and interdependence. In several of the
contexts explored in this study, metaphors of collective navigation
were used to reposition practitioners, learners, and communities as
participants within the same system.

Such narratives do not eliminate differences in authority or
expertise, but they broaden the possibilities for collaboration by
recognising that all participants are responding to the same
structural conditions.


\section{Reflections on the Approach}\label{sec:reflection}

The approach developed in this paper - combining collective
autoethnography, cross-field pattern analysis, and the application of
Reframing Levers - has enabled us to surface dynamics of weaponisation
that are often obscured in everyday professional practice. Although
our cases span multiple domains, the patterns we observed resonate
strongly with challenges familiar in computing: the simplification of
complex socio-technical problems into slogans and metrics, the
individualisation of responsibility for systemic harms, and the
stabilisation of organisational narratives that prioritise efficiency
and innovation over care, maintenance, and justice. While the primary
purpose of our method is generative rather than evaluative, several
observations emerged that shape how we understand both its strengths
and limitations.

Working across diverse professional domains clarified the value of
treating weaponisation not as a series of isolated incidents but as a
patterned social phenomenon. Despite major contextual differences
between education, justice, public administration, research
management, and computing, contributors repeatedly identified similar
mechanisms: simplification, individualisation, moralisation, binary
framing, and organised ignorance. This suggests that the conceptual
vocabulary developed here may have wider applicability than any single
context, while still requiring situated interpretation.

The use of reframing levers proved particularly helpful in supporting
practitioner-oriented reflection. Because the levers are small in
scale and framed as directional shifts rather than prescriptive steps,
co-authors were often able to apply them without requiring formal
authority or institutional permission. This aligns with literature on
micro-resistance and radical incrementalism, which emphasises the
cumulative impact of modest interventions and everyday acts of
sensemaking. At the same time, the examples showed that such levers do
not remove structural constraints; rather, they create openings in
which alternative interpretations and actions become more thinkable.

The collaborative nature of the method, especially the peer annotation
and synthesis stages, also deepened the analysis. Contributors noted
that their own contributions became clearer and more structurally
aware through the process of reading and responding to others. This
kind of mutual recognition and co-interpretation is central to
collective autoethnographic work, but it also echoes broader ideas in
peacebuilding and narrative scholarship: new stories, and new
possibilities for action, often emerge through relational processes
rather than solitary insight.

There are, however, clear limitations. The contributions presented
here are shaped by the particular experiences and positionalities of the authors. The
reframing levers themselves remain provisional: they will likely
evolve as more examples accumulate, and as practitioners in other
fields test, adapt, or contest them. The approach also depends on a
willingness to engage in slow, reflective work, which can be difficult
to sustain in time-poor, performance-driven environments.

Across the examples, the reframing levers did not eliminate the
structural constraints shaping professional practice. What they did
was interrupt narrative closure. By surfacing hidden assumptions,
redistributing responsibility, and making systems visible, the
reframing moves created openings in which different questions
could be asked and different responses considered. For the LIMITS
community, this is particularly relevant. Computing often operates
within narratives that privilege efficiency, optimisation, and
technological solutionism while obscuring social and ecological
consequences. The reframing prompts demonstrated here suggest that
small shifts in framing - toward structural literacy, system insight,
and more generous narratives of professional responsibility - may help
practitioners and researchers notice and challenge these dynamics. In
this sense, reframing becomes not only a reflective practice but also
a practical method for expanding what computing professionals can see,
question, and imagine within socio-technical systems.

\section{Implications for Computing Practice and the LIMITS Community}\label{sec:computing}

Following Vodeb and Escobar's advice to extend our epistemological gaze but to bring lessons back to the discipline with the aim to change it \citep{Vodeb&Escobar}, the insights developed throughout this paper return directly to the concerns of the LIMITS community. The cases and Reframing Levers that emerged from cross-field collaboration provide patterns that resonate strongly with challenges familiar in computing research, education, and professional practice.

First, the phenomenon of weaponisation observed across sectors mirrors long-standing critiques within computing: the simplification of complex socio-technical issues into slogans, metrics, and binary framings; the individualisation of responsibility for systemic harms; and the stabilisation of organisational logics that prioritise efficiency, innovation, and optimisation over care, maintenance, and justice \citep{nardi2018computing,becker2023insolvent, delaBellacasa}. These mechanisms also resonate with wider critiques in computing, HCI, STS, and sociology. The conversion of systemic problems into matters of personal responsibility is visible in critiques of surveillance capitalism, where notice-and-consent regimes shift responsibility to users while leaving corporate incentives and infrastructural asymmetries intact \citep{zuboff2023age}. Similar displacements appear in responses to algorithmic bias, where responsibility for harmful outcomes can be passed to datasets, downstream users, or deployment contexts, even when the design and institutional conditions of the system remain central to the harm \citep{buolamwini2018gender,benjamin2023race}. Likewise, the unrecognised labour that sustains visible institutional or technical commitments has clear precedents in work on invisible labour and ghost work, where the smooth appearance of digital systems depends on work that is rendered marginal, hidden, or poorly valued \citep{gray2019ghost,stilgoe2022ghost}. The cross-domain patterns documented in this paper therefore help illuminate why certain narratives remain so persistent in computing and why efforts toward just, care-centred, or sustainability-oriented practice so often encounter resistance.

Second, the Reframing Levers offer a set of small-scale, practice-oriented interventions that can support more reflective and structurally aware forms of computing. These levers are discursive and relational rather than technical: they invite practitioners, educators, students, and researchers to broaden frames, surface assumptions, resist harmful simplifications, and cultivate more generous and justice-oriented narratives. Such interventions align with LIMITS calls to attend to underlying worldviews and to foreground sufficiency, responsibility, and social-ecological embeddedness \citep{nardi2018computing,becker2023insolvent}.

Third, the cross-field nature of our approach underscores the importance of learning from domains beyond computing. Many of the dynamics described in this paper, such as managerialisation, responsibilisation, and performance-driven cultures, are shared across professional systems. As computing practitioners increasingly work in interprofessional and cross-sector environments, the ability to recognise and gently disrupt harmful narratives becomes a vital competence.

Finally, this work suggests that narrative and framing deserve greater attention within computing curricula, governance structures, and research practices. Stories shape not only how we diagnose problems but also how we imagine the futures computing should help bring about. By foregrounding narrative analysis and reflective reframing as legitimate components of computing practice, the LIMITS community can help cultivate professionals who are better equipped to work within social and ecological limits.

\section{Conclusion}\label{sec:conclusion}

Across sectors and professional contexts, practitioners increasingly confront narratives that narrow responsibility, simplify complex situations, and stabilise harmful organisational logics. In this paper, we have sought to understand these dynamics through the concept of weaponisation and to offer small-scale, peaceful, and progressive interventions in the form of Reframing Levers. Through collective autoethnography, multi-level analysis, and cross-field synthesis, we have shown how reflective reframing can widen perception, restore complexity, and reorient practice toward structural awareness and justice.

Our cases demonstrate that weaponised narratives are not merely rhetorical problems; they materially shape how professionals interpret situations, understand their obligations, and imagine what responses are possible. By identifying shared mechanisms across diverse fields, we have argued that narrative work is an essential component of responding to social and ecological limits—not an optional supplement to technical or managerial expertise.

The Reframing Levers presented here are intentionally modest. They do not claim to solve systemic harms or replace the need for large-scale structural transformation. Instead, they offer practitioners small but meaningful ways to intervene in the stories that organise their own work. These interventions can complement broader institutional, political, and infrastructural efforts while also supporting the cultivation of professional communities capable of imagining and enacting more just and sustainable futures.

As computing continues to grapple with its social and ecological responsibilities, we hope this work contributes to the LIMITS community’s ongoing efforts to foreground care, sufficiency, and justice. By attending to the narratives that shape practice—and by learning from experiences across fields—we can better understand how to support computing within limits and how to challenge the stories that hold the status quo in place.

\bibliographystyle{acm}

\bibliography{references}

\end{document}